# CONCORDANCE OF X-RAY CLUSTER DATA WITH BBN IN MIXED DARK MATTER MODELS

Russell W. Strickland and David N. Schramm[1]

The University of Chicago, Chicago, IL 60637

russell@oddjob.uchicago.edu

dns@oddjob.uchicago.edu

## ABSTRACT

If the hot, X-ray emitting gas in rich clusters forms a fair sample of the universe (as in Cold Dark Matter (CDM) models), and the universe is at the critical density, $\Omega_T = 1$, then the data appears to imply a baryon fraction, $\Omega_{b,x}$ ($\Omega_{b,x} \equiv \Omega_b$ derived from X-ray cluster data), larger than that predicted by Big Bang Nucleosynthesis (BBN). While various other systematic effects such as clumping can lower $\Omega_{b,x}$, in this paper we use an elementary analysis to show that a simple admixture of Hot Dark Matter (HDM, low mass neutrinos) with CDM to yield mixed dark matter shifts $\Omega_{b,x}$ down so that significant overlap with $\Omega_b$ from BBN can occur for $H_0 \lesssim 75$ km/sec/Mpc, even without invoking the possible aforementioned effects. The overlap interval is slightly larger for lower mass neutrinos since fewer cluster on the scale of the hot X-ray gas. We illustrate this result quantitatively in terms of a simple isothermal model. More realistic velocity dispersion profiles, with less centrally-peaked density profiles, imply that fewer neutrinos are trapped and, thus, further increase the interval of overlap. However, we also note that if future observations of light element abundances find that $\Omega_b h^2 \lesssim 0.018$, the range of concordance in this simple mixed dark matter model vanishes.

*Subject headings:* cosmology – neutrinos – dark matter – Hubble constant

---

[1]also NASA/Fermilab Astrophysics group, Fermilab, Batavia, IL 60510

## 1. Introduction

Big Bang Nucleosynthesis (BBN) has repeatedly been found to be in agreement (within the tolerances of the measurements) with observed light element abundances. Along with these observations, BBN predicts a baryon density,

$$0.007h^{-2} \lesssim \Omega_b \lesssim 0.024h^{-2}, \tag{1}$$

(Copi, Schramm, and Turner 1995 a, b, c) where $h \equiv H_0/(100 \text{ km/sec/Mpc})$. Recent work (Barlett *et al.* 1994, White *et al.* 1993) using the ROSAT and ASCA X-ray satellites has demonstrated that rich clusters of galaxies contain significant amounts of hot X-ray emitting gas. If one assumes that the gas is gravitationally confined, then the temperature of the gas can be related to the gravitational potential and, hence, the total mass, $M_T$, of the X-ray emitting region. The intensity of the emission is related to the density of the gas (baryons) in the emitting regions. The mass of the galaxies within the cluster is computed based upon photometric luminosities of the galaxies (e.g., Godwin and Peach 1977) and an empirical mass-luminosity relation (van der Marel 1991). If these clusters are fair samples of the universe, then the ratio of the hot gas mass plus the luminous mass in the galaxies, $M_b$, to the total mass, $M_T$ is a measure of $\Omega_{b,x}$, the value of $\Omega_b$ derived from X-ray cluster data, $(M_b/M_T = \Omega_{b,x}/\Omega_T)$.

White *et al.* (1993) show that the current data (using the simplest "1$^{\text{st}}$ order" analysis) imply

$$0.005 + 0.03h^{-3/2} \lesssim \frac{M_b}{M_T} \lesssim 0.013 + 0.07h^{-3/2}, \tag{2}$$

where first term refers to the galaxies and the second term refers to the X-ray gas. Fig. 1 (adapted from Gott *et al.* 1974, Copi and Schramm 1995) shows the comparison of $\Omega_b$ from BBN (Eqn. 1) and $\Omega_{b,x}$ from $M_b/M_T$ (Eqn. 2) assuming that the clustered gas is representative of the baryon density in an $\Omega_T = 1$ universe. Notice that there is only overlap for $H_0 \lesssim 55$ km/sec/Mpc. Of course, if $\Omega_T < 1$ or if the clusters are not a fair sample, then there is no conflict for larger values of $H_0$. Furthermore, systematics (c.f. Bird, Mushotzky, and Metzler 1995) seem to be capable of shifting $M_b/M_T$ primarily to lower, not higher, values. For example, density inhomogeneities in the hot gas would produce an anomolously high implied gas mass. Also, if significant magnetic fields were present in the cluster such that the magnetic pressure was non-negligible compared to the thermal pressure, then $M_T$ would be larger than the value derived by considering the thermal pressure alone. Furthermore, weak lensing estimates of $M_T$ (Squires *et al.* 1995) seem to yield slightly larger values than those from the X-ray data, which could further lower $M_b/M_T$. In this paper, however, we will accept the $M_b/M_T$ taken from the simplest "1$^{\text{st}}$ order" analysis of the X-ray data but will argue that a Hot Dark Matter (HDM) component (neutrinos) can allow for a greater overlap. Obviously in a pure HDM model with structure formed by the wakes of cosmic strings (Brandenburger 1994, Stebbins 1995) clusters are unlikely to be fair samples. However, in this paper we will show that even in standard, inflation-inspired models an HDM admixture can increase the interval of overlap.

In pure Cold Dark Matter (CDM) models, the scale of clusters ($R \gtrsim$Mpc) have obtained a fair sample of the universe and, hence,

$$\Omega_{b,x} = \Omega_T M_b/M_T. \tag{3}$$

Since the most interesting case is $\Omega_T = 1$, we will focus our attention on this option. When one switches to a mixed dark matter model, with HDM as well as CDM, $\Omega_T$ in Eqn. 3 becomes the cosmological density of all of the matter which clumps on the scale of the cluster, $\omega_{cluster} \equiv b\Omega_{cluster}$, where $\Omega_{cluster} = 0.2 \pm 0.1$ (Faber and Gallagher 1979) and $b \sim few$ is a bias factor which need not be the same as the galaxy formation bias factor used in some CDM models of structure formation. Thus,

$$\omega_{cluster} = \Omega_b + \Omega_{CDM} + f \cdot \Omega_{HDM}, \tag{4}$$

where $f$ is the fraction of HDM which clumps within the cluster. Mixed dark matter (HDM + CDM) models (Occhinero 1985) have been shown to be a viable way to enable COBE-normalized CDM-like models to fit galaxy correlation data, for $\Omega_{CDM} = 0.8$ with $\Omega_T = 1$ (Primack *et al.* 1995, Davis 1995, Ma and Bertschinger 1994).

## 2. Low Mass Neutrinos

Neutrinos are the best candidate for the HDM. The sum of the masses of all of the species of neutrinos (and anti-neutrinos) yields (Lee and Weinberg 1977, Cowsik and McClelland 1972, Marx and Szalay 1972)

$$\Omega_{HDM} = \frac{\sum_i m_{\nu_i}}{92 h^2 \text{ eV}}. \tag{5}$$

The fraction of neutrinos which will contribute to the cluster is $f = \varepsilon_{HDM}/\varepsilon_{(b\,\&\,CDM)}$, where $\varepsilon_x$ is the ratio of the density of component $x$ within the cluster to the cosmological density of the same component. In the spirit of the Thomas–Fermi corrections used in models of degenerate electrons within a Coulomb potential (Salpeter 1961), we can write $\varepsilon_{HDM}$ for a single neutrino mass (which could be degenerate and shared by multiple neutrino species) locally as a ratio of Fermi integrals, $\varepsilon_{HDM} = I(m_\nu \phi(r)/kT_\nu)/I(0)$, where $\phi(r) = -GM(r)/r$, $T_\nu = 1.9$ K, and

$$I(\xi) = \int_0^\infty y^2 \frac{\mathrm{d}y}{\mathrm{e}^{(y+\xi)} + 1} \tag{6}$$

is the relativistic Fermi distribution function. (A relativistic distribution is employed since, due to the low cross section for neutrino interactions, the distribution has not changed since the neutrino temperature exceeded about 1 MeV (Lee and Weinberg 1977, Tremaine and Gunn 1979).) We can generalize to neutrino species with different masses by replacing $f \cdot \Omega_{HDM}$ in Eqn. 4 with $\sum_j f_j \cdot \Omega_{HDM(j)}$, where $f_j = f(m_{\nu_j})$ and $\Omega_{HDM(j)} = 2 m_{\nu_j}/(92 h^2 \text{ eV})$, with the factor of 2 arising from summing over $\nu_j$ and $\bar{\nu}_j$. For the remainder of this paper we will consider only the case of a single, possibly degenerate, neutrino mass, as this will serve to suitably illustrate the features of this model.

We defined $\varepsilon$ such that $\chi_{cluster} = \varepsilon_{(b \& CDM)}(\Omega_b + \Omega_{CDM}) + \varepsilon_{HDM}\Omega_{HDM}$, where $\chi_{cluster} \equiv \rho_{cluster}/\rho_{crit}$. Thus,

$$f = (\Omega_b + \Omega_{CDM})\frac{\varepsilon_{HDM}/\chi_{cluster}}{1 - \Omega_{HDM} \cdot (\varepsilon_{HDM}/\chi_{cluster})} \qquad (7)$$

where $\chi_{Coma} \approx 280$ (e.g., White et al. 1993). If $0 \leq \varepsilon_{HDM} \leq \chi_{cluster}$ as one would expect, then $0 \leq f \leq 1$. Fig. 2 shows how $f$ depends on the strength of the cluster potential, $\langle v^2 \rangle$, with the cluster radius fixed at $1.5h^{-1}$ Mpc, the standard Abell radius. Fig. 3 gives the fraction of trapped neutrinos as a function of the radius of the cluster with the potential strength fixed at $(1800 \text{ km s}^{-1})^2$, the value adopted by White et al. (1993). The cluster is parameterized in terms of these two quantities ($\langle v^2 \rangle$ and $R$). Notice from Eqn. 6 that $\varepsilon_{HDM}$ is a function of $\langle v^2 \rangle = \phi \propto M/R$ but not $R$ alone and that $\chi_{cluster}$ is a function of $\langle v^2 \rangle$ and $R$; $\chi_{cluster} \propto M/R^3 \propto \langle v^2 \rangle/R^2$. Thus, Fig. 2 illustrates the effects of changing $\varepsilon_{HDM}$ and $\chi_{cluster}$ since $\langle v^2 \rangle$ is an independent variable in both parameters. However, in Fig. 3 only $\chi_{cluster}$ varies since $\varepsilon_{HDM}$ is not a function of $R$ alone. Fig. 4 gives $f$ as a function of $m_\nu$. Note that $f$ increases with $m_\nu$ since more massive neutrinos are more likely to clump on the scale of rich clusters.

In computing $\varepsilon_{HDM}$ we adopt an isothermal velocity dispersion; therefore, $\varepsilon_{HDM}$ is constant throughout the cluster. This assumption is quite good for our purposes, since even fairly large variations in the actual mass distribution of the cluster have little effect upon the derived fraction of trapped neutrinos. See Fig. 2. Furthermore, an isothermal model is the most conservative choice in that it over-estimates the number of neutrinos bound to the cluster. In an isothermal model, $\rho \propto r^{-2}$. Any model in which the total density profile is less centrally peaked (Bird, Mushotzky, and Metzler 1995, Girardi et al. 1995) will have $|\phi(r)| \leq |\phi(R)|$ everywhere within the cluster. Thus, $f$ will be smaller than in the isothermal case. (See Eqns. 6 & 7.)

Fig. 5 shows the overlap in $\Omega_b$ as derived from BBN (Eqn. 1) and our mixed dark matter interpretation of the Coma cluster data (Eqns. 2, 3, & 4) for $\Omega_{CDM} = 0.8$. It is clear that this overlap increases with decreasing neutrino mass, since fewer neutrinos will clump on the scale of the cluster. However, $\Omega_T = 1$, $\Omega_{b,x}$ is fairly insensitive to $m_\nu$ for a particular value of $\Omega_{CDM}$. To see this, we can expand $\Omega_{b,x}$ in powers of $f$; from Eqns. 3 & 4 we have

$$\Omega_{b,x} \propto 1 + \frac{\Omega_{HDM}}{\Omega_b + \Omega_{CDM}} f + \ldots \sim 1 + 0.2f, \tag{8}$$

for $\Omega_{CDM} = 0.8$ and $h = 0.5$. For this value of $h$, Eqn. 5 specifies a maximum neutrino mass (one $\nu\bar{\nu}$-pair) of $m_\nu \sim 2$ eV. Fig. 4 shows that $f \lesssim 0.005$ for this range of neutrino masses. Thus, changing the neutrino mass, or similarly the number of massive neutrino species, will result in a change in $\Omega_{b,x}$ on the order of several hundredths of a percent.

Although we have shown that the X-ray cluster data can be made to be consistent with the predictions of Big Bang Nucleosynthesis for $H_0 \lesssim 75$ km/s/Mpc, we note that this concordance depends sensitively upon the upper limit of $\Omega_b h^2$ derived from the observed light element abundances. If future measurements of light element abundances find that $\Omega_b h^2$ lies near the present upper limit ($\Omega_b h^2 \approx 0.024$), then the consistency of X-ray cluster data and BBN will be retained. However, if $\Omega_b h^2$ is found to be less than the value representing the intersection of the lower $\Omega_{b,x}$ curve and the line $h = 0.38$ ($\Omega_b h^2 \approx 0.018$), there will no longer be a range of concordance.

The maximum value of $H_0 = 100\,h$ km/s/Mpc consistent with BBN and these cluster data is given by

$$h = \left(\frac{0.02}{0.005 h^{3/2} + 0.03}\right)^2 \left[\Omega_b + \Omega_{CDM} + f\Omega_{HDM}\right]^{-2}. \tag{9}$$

Fig. 6 illustrates the upper limit of the range of consistency for $h$ as a function of $\Omega_{CDM}$ for these observations for a range of $\Omega_{CDM}$ and for various numbers of massive neutrino species. Since $m_\nu \propto \Omega_{HDM}$, the limiting value of $h$ becomes insensitive to the multiplicity of the mass degeneracy for larger values of $\Omega_{CDM}$, where $\Omega_{HDM}$ tends toward zero. This figure also illustrates this decreasing importance of the actual number of massive neutrino species with increasing $\Omega_{CDM}$.

## 3. Conclusions

Structure formation in a Cold + Hot Dark Matter model is initiated by perturbations in the CDM field. The neutrinos are trapped in the primordial fluctuations and thereby accrete onto the new structure. Our analytical model illustrates that neutrinos contribute to the cluster halo. Although the model is simplistic, its results are believable. We find that neutrinos are only a small component of the mass of the cluster. Since the neutrinos are such a small component of the mass of the cluster, $\Omega_T$ is effectively represented by the (smaller) $\omega_{cluster}$ as described in the text. Thus, we have shown that in the framework of a cold + hot dark matter model with massive neutrinos these data are consistent with BBN, $\Omega_T = 1$, and $H_0 \lesssim 75$ km/sec/Mpc. In such a mixed dark matter model the more "radical" options of Fukgita, Hogan, and Peebles (1993) for resolving this issue become unnecessary. As long as observed light element abundances allow $\Omega_b h^2 \gtrsim 0.018$, there will be a range of concordance between X-ray cluster data and BBN. This critical value of $\Omega_b h^2$ will increase if low values of $H_0$ are ruled out. However, as mentioned earlier, numerous systematic effects could shift the cluster data, lowering this critical value and increasing the range of concordance toward larger values of $H_0$.

We acknowledge useful discussions with Mike Turner, Craig Copi, and Evalyn Gates. While this work was in the final stages of preparation we received a preprint from Kofman *et al.* (1995) in which quantitatively similar results were obtained. However, their paper de-emphasized the increased overlap due to their focus on higher values of $H_0$. This work was supported in part by NSF, NASA, and DOE at The University of Chicago and by NASA at Fermilab.

Fig. 1.— Values of $\Omega_b$ allowed by Big Bang Nucleosynthesis and implied by X-ray cluster observations for a CDM model with $\Omega_T = 1$. Concordance between the two methods of determining $\Omega_b$ requires that $H_0 \lesssim 55$ km/sec/Mpc. Also shown are bounds based upon the age of globular clusters ($t_0 > 10$ Gyr) and Type Ia SN determinations of $H_0$ ($H_0 \geq 38$).

Fig. 2.— The fraction of trapped neutrinos as a function of virial velocity dispersion (normalized to the value for the Coma cluster). We have assumed $\Omega_{CDM} = 0.8$ and $\Omega_b = 0.04$ (consistent with BBN for $H_0 \approx 50$ km/sec/Mpc). Thus the sum of all neutrino and anti-neutrino masses is approximately $18h^2$ eV.

Fig. 3.— The fraction of trapped neutrinos as a function of cluster radius (normalized to the radius for the Coma cluster) for a fixed $\langle v^2 \rangle = \langle v^2 \rangle_{Coma}$. See Fig. 2 for details concerning $\Omega_{CDM}$, $\Omega_b$, and $m_\nu$.

Fig. 4.— The fraction of neutrinos of a given mass (in eV) which are trapped within the Coma cluster for an $\Omega_{CDM} = 0.8$, $\Omega_b = 0.04$ model. Note that an $\Omega_{HDM} \approx 0.2$ universe will not allow a $\nu\bar{\nu}$-pair with individual masses in excess of $\sim 5.5$ eV for $h \lesssim 0.8$. Thus, $m_\nu \gtrsim 5.5$ eV requires that $\Omega_{CDM} \lesssim 0.8$.

Fig. 5.— Values of $\Omega_b$ concordant with Big Bang Nucleosynthesis and X-ray cluster observations. The cluster model assumes that $\Omega_{CDM} = 0.8$, $\Omega_T = 1$, $h = 0.5$, and $m_\nu = 2$ eV, consistent with a single massive $\nu\bar{\nu}$-pair. Values of $\Omega_b$ consistent with both methods require $H_0 \lesssim 75$ km/s/Mpc. See Fig. 1 for details about the other features of this plot.

Fig. 6.— Maximum values of $h$ from concordance of Big Bang Nucleosynthesis and "1$^{\text{st}}$ order" X-ray cluster observations as a function of $\Omega_{CDM}$. Note that these curves do not represent particular values of $m_\nu$ since the total mass in neutrinos is a function of $h$ and $\Omega_{HDM} = 1 - (\Omega_b + \Omega_{CDM})$. The inset shows the dependence of the concordance interval upon the number of massive neutrino species. The three curves converge for larger values of $\Omega_{CDM}$ since $m_\nu \propto \Omega_{HDM}$ tends toward zero. Thus, for reasonable values of $\Omega_{CDM}$, the number of massive neutrino species is unimportant, provided that there is at least one.